\documentclass{article}

\usepackage{arxiv}

\usepackage[utf8]{inputenc} 
\usepackage[T1]{fontenc}    
\usepackage{hyperref}       
\usepackage{url}            
\usepackage{booktabs}       
\usepackage{amsfonts}       
\usepackage{nicefrac}       
\usepackage{microtype}      
\usepackage{lipsum}		
\usepackage{graphicx}
\usepackage{natbib}
\usepackage{doi}
\usepackage{amsmath}

\title{\textbf{Sovereign AI: Rethinking Autonomy in the Age of Global Interdependence} }

\author{ \href{https://orcid.org/0009-0004-8549-5168}{\includegraphics[scale=0.06]{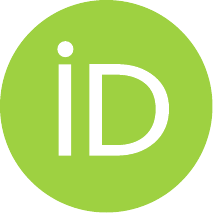}\hspace{1mm}Shalabh Kumar Singh}\thanks{Address for correspondence: 9th Floor, Block 14, Pritech SEZ, Outer Ring Road, Bellandur Village, Bengaluru 560103, Karnataka, India.} \\
	\textsuperscript{1}Principal Director - Technology Thought Leadership,\\
    Accenture Research, \\
    Growth and Strategy\\
    Bengaluru\\
	\texttt{shalabh.kumar.singh@accenture.com} \\
	\And
	\href{https://orcid.org/0000-0000-0000-0000}{\includegraphics[scale=0.06]{orcid.pdf}\hspace{1mm}Shubhashis Sengupta} \\
	Managing Director, \\
    Accenture Innovation, \\
    Accenture Innovation Hub\\
    Bengaluru\\
	\texttt{shubhashis.sengupta@accenture.com} \\
}

\renewcommand{\shorttitle}{\textit{Sovereign AI}}

\hypersetup{
pdftitle={\textbf{Sovereign AI: Rethinking Autonomy in the Age of Global Interdependence} },
pdfsubject={q-bio.NC, q-bio.QM},
pdfauthor={Shalabh Kumar Singh, Shubhashis Sengupta},
pdfkeywords={First keyword, Second keyword, More},
}

\begin{document}
\maketitle

\begin{abstract}
\ Artificial intelligence (AI) is emerging as a foundational general-purpose technology, raising new dilemmas of sovereignty in an interconnected world. While governments seek greater control over it, the very foundations of AI—global data pipelines, semiconductor supply chains, open-source ecosystems, and international standards—resist enclosure. This paper develops a conceptual and formal framework for understanding sovereign AI as a continuum rather than a binary condition, balancing autonomy with interdependence. Drawing on classical theories, historical analogies, and contemporary debates on networked autonomy, we present a planner’s model that identifies two policy heuristics: equalizing marginal returns across the four sovereignty pillars and setting openness where global benefits equal exposure risks.

We apply the model to India, highlighting sovereign footholds in data, compute, and norms but weaker model autonomy. The near‑term challenge is integration via coupled Data×Compute investment, lifecycle governance (ModelOps), and safeguarded procurement. We then apply the model to the Middle East (Saudi Arabia and the UAE), where large public investment in Arabic‑first models and sovereign cloud implies high sovereignty weights, lower effective fiscal constraints, and strong Data×Compute complementarities. An interior openness setting with guardrails emerges as optimal. Across contexts, the lesson is that sovereignty in AI needs managed interdependence, not isolation. 
\end{abstract}


\textit{\textbf{Disclaimer}}: This article reflects the personal views of the authors and does not represent the positions or policies of their employers or affiliated institutions.  

\section{Introduction}
\ With artificial intelligence (AI) rapidly becoming the foundation for economic, military, and civic life, many nation-states are asserting the need for sovereign control—advocating greater control over its development and deployment. Yet the very nature of AI resists enclosure. As with earlier infrastructures such as electricity (Hughes, 1983) and the internet (Castells, 1996), AI is raising profound questions of sovereignty. Who controls the data, the compute infrastructure, the models, and the norms that govern this technology? For many states, especially in the Global South, these questions are not abstract—they touch on cultural identity, strategic autonomy, and institutional legitimacy.

The most advanced foundational models today are trained on globally sourced data, rely on transnational cloud infrastructure and global supply chains, and evolve through collaborative open-source ecosystems. This raises a fundamental dilemma: attempts to implement sovereign AI policies risk undermining the innovation and interoperability that make these systems viable in the first place.

Governments increasingly worry that dependence on foreign providers for critical AI infrastructure undermines national interest. Currently, large US based private players dominate the global AI infrastructure, foundation models, data and benchmarks. In response, we see homegrown approaches, such as Mistral, which is positioning itself as a public funded, open-source European alternative (Sam Schechner, 2025). The European Commission’s (2025) State of the Digital Decade report calls for targeted public and private investment in AI infrastructure, connectivity, and talent to ensure technological sovereignty and reduce dependence on external providers for AI and cloud services. Saudi Arabia has launched its own state-backed AI company, HUMAIN, to develop multimodal Arabic large language models for regional and global markets (Fatima, 2025). UAE has partnered with Microsoft and Core42 with the ambition to become the world’s first fully AI-native government by 2027, underpinned by a sovereign cloud.

India too has entered the fray with the Rs. 10,000 crore (approximately US \$ 1.14 bn) IndiaAI Mission, which combines state-backed investment in national compute infrastructure with the development of open, multilingual AI tools by private firms tailored to the Indian context—such as Sarvam AI (India’s first sovereign large language model) and Bhashini (AI powered language translation platform for making digital content and services available to India’s diverse linguistic population). 

However, the question of AI sovereignty, like that of other general-purpose technologies before it, raises complex dilemmas. What does it truly mean for AI to be sovereign? Is it simply about national control over AI infrastructure and software stack? Does it align with the vision of technological self-sufficiency?

This paper makes three contributions. First, it situates AI sovereignty within broader theories of sovereignty and draws historical lessons from foundational technologies. Second, it defines sovereign AI through four pillars—data, compute, models, and norms—while situating them within governance and interdependence debates. Third, it introduces a formal planning model of sovereignty–openness trade-offs and applies it to India and the Middle East, offering lessons for other economies. The central argument is that AI sovereignty is not binary but a continuum, shaped by how countries allocate resources, build institutions, and engage internationally. 
\

\section{Sovereignty and Technology: Theory and History}
\label{sec:headings}
Political theory provides useful lenses for thinking about AI sovereignty. Hobbes viewed sovereignty as an indivisible authority (state) necessary for social order. This could mean a state-controlled AI system. Locke advocated for limited and accountable rule based on a social contract between the governed and the governing, which would suggest AI regulated for national purpose. Rousseau introduced the notion of the general will, with sovereignty reflecting the expression of the collective. This would mean a decentralized approach allowing a combination of stakeholders—governments, businesses, and civil society—to play a role in its governance.

Held views states and societies as being “enmeshed in webs of international conditions and processes as never before”, suggesting a shared, globally accountable, and democratically governed control over AI's development and deployment. From Held’s perspective, national "Sovereign AI" initiatives are perhaps a potentially problematic response, but they could be a stepping stone to a cosmopolitan approach to governance. 

Carl Schmitt (1922) in \textit{Political Theology} argues that sovereignty is the power to declare the exception—to determine when rules can be suspended in times of crisis. This raises another important question. When AI is involved in high-stakes decision making—such as affecting national security, financial markets or economic policy—who has the authority to override its actions? Does it require an independent regulator, or can governments and private sector make these decisions?

In short, sovereign AI cannot simply be defined in terms of restricting foreign technology. It is a deeper question of strategic intent and control, where nations must choose which components of AI to control for strategic sovereignty and which dependencies to accept.

The philosopher Antonio Gramsci’s (1971) concept of cultural hegemony offers another insight—sovereignty is not merely controlling resources but maintaining control over the dominant narratives and ideology. Even if AI is developed within the national boundaries but is trained on datasets that reflect values, norms, political ideology, and economic systems of other countries, would that compromise its sovereignty? Seen thus, power is exercised not only through resources but by controlling dominant narratives, including linguistic and cultural hegemony. In defining sovereignty, then, who gets to decide what is acceptable and what is not—especially in a country like India, which is both culturally diverse and grappling with the desire to slow the westernization of its social and moral frameworks? Given that value systems differ across cultures, a piece of advice from an AI model that may be socially acceptable in one context could be inappropriate in another. This raises not only technical and institutional questions, but normative ones: whose values, languages, and ethics are embedded in sovereign AI systems, and with what degree of democratic legitimacy?

Ultimately, only governments hold the formal authority to enforce AI sovereignty. In a federal democracy like India, this raises additional complexity: while the central government owns the AI infrastructure and policy, normative values like cultural acceptability and linguistic inclusion often fall within the purview of state governments and regional institutions. Defining sovereign AI, therefore, requires recognition of India’s federal character and social pluralism. Governments—both national and subnational—will need to decide whether they want to nationalize AI development, strictly regulate private-sector AI firms, or allow flexibility for corporations to drive AI progress while operating within a regulatory framework.

The conventional state-centric model of sovereignty is, however, challenged by Swati Srivastava and Justin Bullock (2024), who argue that AI governance will need to be shaped by negotiations between governments, corporations, and international institutions. Sovereign AI is not the first foundational technology to face this concern. Electricity and the internet were once considered state-controlled sovereign resources. Early electricity grids were usually state-controlled, and the internet started as the government-funded ARPANET in the U.S. But as Thomas Hughes (1983) shows, the growing scale and complexity of electrical systems—driven by economies of scale, engineering coordination, and institutional adaptation—led to regional and international energy networks that extended beyond single-state control. And the internet evolved into a decentralized, globally connected network of communication and information exchange, where no single nation-state has complete control (Castells, 1996). The same could happen to AI, which requires vast computational resources, specialized infrastructure, diverse training data, and cross-border innovation. 

For developing countries like India, this poses both a constraint and a strategic opening. India’s approach towards sovereign AI reflects a strategy of managed interdependence—localizing sensitive infrastructure and promoting linguistic and cultural inclusion, while remaining engaged with transnational AI ecosystems. The IndiaAI Mission—supported by public infrastructure such as AIRAWAT, Aadhaar, and UPI—seeks to assert policy autonomy by building foundational capacity. However, significant limitations remain: dependence on global semiconductor supply chains, limited AI R\&D investment, a skills gap in advanced AI development, and a still-evolving regulatory framework.

\section{Interdependence as a key feature of sovereignty}
\label{sec:headings}
The Complex Interdependence Theory of Keohane and Nye (1977) argues that sovereignty in the modern era is about managing strategic dependence. No single nation can achieve complete AI sovereignty given that AI value chains are globally interconnected. The United States, the leader in AI development, sources critical minerals from overseas partners for chip production. China relies on semiconductor technology from the US in developing its AI ecosystem. The EU relies on US-based cloud providers. 

This means that sovereign AI reflects a networked autonomy not isolation, as Srivastava \& Bullock (2024) argue. Like strategic autonomy in foreign policy, networked autonomy recognizes that complete self-reliance in a globally interdependent AI ecosystem is neither feasible nor desirable. Nations need to selectively engage in international AI partnerships, open-source collaborations, and cross-border regulatory agreements, while exercising control over key AI components, such as data governance, model alignment, compute infrastructure, and regulatory norms. For example, nations may contribute to open-source foundation models but require local security protocols for inference infrastructure, underpinned by hybrid logic of trusted cloud” frameworks, digital commons, and federated AI governance.

The trade-off between security and innovation remains core to this discussion. International collaborations, open-source research communities, and cross-border knowledge exchange have been critical to AI breakthroughs. The European Union’s AI strategy clearly shows this duality—retaining regulatory autonomy but promoting collaborative research among member states. Any pragmatic approach to sovereign AI has to embrace this reality of controlled interdependence rather than strict isolation. 

India, as a founding member of the Global Partnership on Artificial Intelligence (GPAI), has signalled an intent to shape global norms from within, embodying the idea of networked autonomy in practice. Yet its ability to govern interdependence is limited by the absence of institutional frameworks to influence international technical standards and limited domestic capacity in advanced chip design. 

For developing economies like India, building multi-stakeholder institutions—involving civil society, academia, and the private sector—will be essential to engage in “norm entrepreneurship” (Finnemore and Sikkink, 1998) to actively shape international norms aligned with its own democratic and developmental priorities. These institutions can serve as the foundation for sustained participation in global standard-setting bodies and for conducting coordinated technical diplomacy.

\section{Implementing Sovereign AI: Technical Dimensions}
\label{sec:headings}

Technically, sovereignty can be implemented at either the training or the inference stages, or both. Training requires vast amounts of data to improve accuracy and precision while reducing bias. This means any strict definition of sovereign AI would necessitate control over data pipelines. 

However, it is important to distinguish data localization from data sovereignty. Data localization refers to legal or regulatory requirements that data generated within a country is stored locally, but it does not preclude foreign entities from accessing and managing it, say for federated learning (Chander and Lê 2015). In contrast, data sovereignty implies full jurisdictional and infrastructural control—covering not only data but also the technology stack that stores and processes it. European Union’s GAIA-X initiative illustrates this attempt to secure this deeper level of autonomy through trusted infrastructure and governance frameworks (Federal Ministry for Economic Affairs and Energy 2025).

True AI sovereignty involves more than geographic control over data. It includes ownership of data centers, compute infrastructure, chip fabrication, control over APIs and inference protocols, and alignment of model behaviour with national legal, ethical, and cultural standards. Merely complying with localization mandates may offer regulatory oversight, but it does not remove upstream dependencies—such as foreign cloud services, foundational APIs, or pre-trained models—that can shape the system’s functioning. 

At the same time, excessively rigid data localization measures hinder innovation and impose economic costs by restricting access to multilingual, multicultural, and domain-diverse datasets. Maximalist data sovereignty strategies—such as banning foreign cloud providers—can undermine innovation, competitiveness, and digital trade, especially for smaller economies (Cory and Dascoli 2021). A more pragmatic vision of sovereign AI must therefore balance autonomy with selective openness and interoperability.

Compute sovereignty is another challenge—most advanced AI chips are designed by U.S. firms and manufactured in Taiwan and South Korea. Without domestic semiconductor production, AI sovereignty remains vulnerable to global supply chain disruptions (Ebrahimi, 2024).

Inference raises other concerns. For example, can inference be run on foreign cloud infrastructure? It also introduces the risk of API dependencies—reliance on foreign APIs for speech recognition, translation, or other tasks, can potentially compromise autonomy. On-premises deployments or secure national cloud infrastructures may come at significant cost. 

Another issue is post-training ModelOps—the operations required to manage machine learning models through their lifecycle. Even if an AI model is initially sovereign, continuous updates, reinforcement learning from user feedback (RLHF), and integration with external services could gradually introduce foreign dependencies. In the absence of robust retraining cycles, monitoring tools, and update governance, models may gradually see behavioural shifts that compromise the original intent. AI sovereignty is not just a one-time achievement but an ongoing process of maintaining control over how models evolve through long-term institutional capacity to audit and adapt models.

As evidenced in India’s current approach, it risks building sovereign inputs—such as a roadmap for public compute infrastructure and indigenous model development—without securing sovereign outcomes. Beyond its structural dependencies in chip design, inference cloud infrastructure, and secure data curation, India also lacks a coordinated approach to ModelOps, limiting its ability to govern model updates, bias correction, and post deployment behaviour. Without governance mechanisms across the AI lifecycle, sustained institutional control over how models behave, evolve, and are held accountable can be a challenge.

\section{\textbf{A Planning Model for Managed Interdependence}}

As is evident from the discussion above, Sovereign AI is not a fixed, binary concept but a dynamic process of adaptation and control. Some dependencies are inevitable and acceptable while others can be controlled. As with technologies like electricity and the internet, sovereign AI may end up being defined by governance standards developed through alliances of nations.

Sovereignty in AI may also be asserted in diverse ways. Countries like the UAE and Saudi Arabia are pursuing state-led models of AI sovereignty. At the other end of the spectrum, it may be a market-driven approach, led by private firms with national support, which is closely approximated by the US. EU’s GAIA-X initiative offers a hybrid, federated approach, grounded in open-source principles and regional trust that pools public and private actors.

The challenges outlined in the sections above, both conceptual and technical, show that sovereignty cannot be a one-off achievement: it requires ongoing institutional and technical capacity. 

In this context, we formalize AI sovereignty as a planner’s problem of managing interdependence, which is likely to be relevant for developing economies like India. We define Sovereign AI across four interdependent pillars:

\begin{itemize}
    \item \textbf{Data (D)} – Ownership, governance, and curation of training and operational datasets. Data localization laws are one expression, but true data sovereignty implies exclusive domestic control over collection, storage, and usage pipelines.
\end{itemize}

\begin{itemize}
    \item \textbf{Compute Infrastructure (C)} – This includes chips, servers, and sovereign cloud infrastructure.
\end{itemize}

\begin{itemize}
    \item \textbf{Model Autonomy (M)} – Control over large language models and foundation models.
\end{itemize}

\begin{itemize}
    \item \textbf{Normative Alignment (N)} – Embedding local languages, values, and cultural priorities.
\end{itemize}

We define \textbf{AI Sovereignty (S)} as a function of these four pillars:

\begin{equation} \ S = f (D, C, M, N)\end{equation}

These four pillars are not only integral to how sovereignty would play out in practice, but also the levers that policy makers can actually move.

Let a country choose policy investments on each of these four pillars, denoted by  \(x_i\), 
bounded by a public budget or PPP funds:

\begin{equation}
x \equiv (x_D,x_C,x_M,x_N )\le B, 
\end{equation}  \;  where B is the total public budget.

Let  $\mu \ge 0$ reflect the shadow price of public funds.

We translate spending into normalized capacities in [0, 1] using saturating exponentials (for diminishing returns and interpretability):

\begin{equation}
D(x_D) = 1 - e^{-a_D x_D}, \end{equation}
\begin{equation} C(x_C) = 1 - e^{-a_C x_C}, \end{equation} 
\begin{equation} N(x_N) = 1 - e^{-a_N x_N},
\end{equation}

where $a_i>0$ are absorptive‐capacity/productivity parameters (higher $a_i$ means every dollar delivers more capacity--better institutions, supplier depth, execution).

\paragraph{Model autonomy with complementarity.}
Model control depends not only on the direct model spend but also the complementarity strength between data and compute infrastructure. It is denoted as: 
\begin{equation}
M(x_M; D,C) \;=\; \min\!\Big\{\,1,\;\; 1 - e^{-a_M x_M} + \theta\,D(x_D)\,C(x_C)\,\Big\},
\end{equation}
with $\theta\ge 0$ measuring complementarity. If either $D$ or $C$ lags, model control $M$ is hard to achieve in practice .

\paragraph{Sovereignty index and openness.}
Define a sovereignty index as a weighted sum:
\begin{equation}
S(x) \;=\; w_D D + w_C C + w_M M + w_N N,
\quad w_i\ge 0,\;\; \sum_i w_i = 1.
\end{equation}  
$w_i$ represent policy weights reflecting priorities, such as value placed on normative alignment. 

Openness $O\in[0,1]$ captures international participation, such as joint participation, open models, cross-border APIs/cloud, and joint research (where 0 = autarky, 1 = highly networked). It brings benefits (e.g., spillovers, scale, talent) but raises exposure or dependency risks. We model this simply as follows.
\begin{itemize}
    \item $G(O)= g\,\ln\!\bigl(1 + k\,O\bigr)$ measures the benefits with diminishing returns to increasing openness. Here, k is a parameter that scales the benefits from openness. 
    \item $P(O)= p\,O$ assumes a linear exposure cost for simplicity, although it can be converted into a convex function. In this, p represents the marginal cost of exposure. 
\end{itemize}

\paragraph{Planner’s problem.}
The problem statement for the policy makers is to maximize the welfare function subject to the budget constraint and choose $x\ge 0$ and $O\in[0,1]$:
\begin{equation}
\max_{x\ge 0,\; 0\le O\le 1}\;
W \;=\; \alpha\,S(x) \;+\; (1-\alpha)\,G(O) \;-\; \lambda\,P(O)
\quad\text{s.t.}\quad
x_D + x_C + x_M + x_N \;\le\; B,
\end{equation}

where $\alpha\in[0,1]$ is the policy weight on sovereignty vs.\ openness, and $\lambda\ge 0$ reflects the sensitivity to risk.

\paragraph{{Lagrangian.}}
With $\mu\ge 0$ as the multiplier on the budget constraint, the Lagrangian is
\begin{equation}
\mathcal{L}
=\alpha S(x) + (1-\alpha)G(O) - \lambda P(O)
+ \mu\Big(B - x_D - x_C - x_M - x_N\Big).
\end{equation}
\paragraph{First-order conditions (FOCs).}

\begin{itemize}
    \item For spending on each pillar:
\end{itemize}

\begin{equation}
\frac{\partial\mathcal{L}}{\partial x_i}=\alpha \, \frac{\partial S}{\partial x_i} \;- \; \mu \; = \; 0 \end{equation}

Data:
\[
\alpha \frac{\partial S}{\partial x_D} = \mu
\]
\[
\implies \alpha \cdot (w_D a_D e^{-a_D x_D} + w_M \theta\, C(x_C)\, a_D e^{-a_D x_D}) = \mu
\]

Compute: \[
\alpha \frac{\partial S}{\partial x_C} = \mu \]
\[\implies \alpha \cdot (w_C a_C e^{-a_C x_C} + w_M \theta D(x_D) a_C e^{-a_C x_C}) = \mu
\] Models: \[
\alpha \frac{\partial S}{\partial x_M} = \mu \]
\[\implies \alpha \cdot (w_M a_M e^{-a_M x_M}) = \mu \]
Norms:
\[
\alpha \frac{\partial S}{\partial x_N} = \mu \]
\[\implies \alpha \cdot (w_N a_N e^{-a_N x_N}) = \mu \]

\begin{itemize}
    \item For openness $O$:
\end{itemize}

\begin{equation}
\frac{\partial\mathcal{L}}{\partial O}=(1 - \alpha) \, G'(O) \;- \; \lambda P'(O) \; = \; 0 \end{equation}
Substituting functional forms,

\begin{equation}
    (1-\alpha) \frac{gk}{1+kO} - \lambda p \, - 1 \, = \, 0 
\end{equation}

\begin{equation}
    O^* \, = \, \frac{(1 -\alpha) g}{\lambda P} - \frac {1}{k}  
\end{equation}
Since $O\in[0,1]$, the true optimum is
\begin{equation}
 O^* = \max\left\{0,\min\left[1,\frac{(1-\alpha)g}{\lambda p}-\frac{1}{k}\right]\right\}
\end{equation}
\clearpage
 {Implications for policymakers:}

\begin{itemize}
    \item The equilibrium level $O^*$ shows that complete autarky or full openness is rarely the best option. Instead, the best option for most countries would be managed interdependence.
\end{itemize}
\begin{itemize}
    \item Policy makers must allocate their budget in such a way that each pillar (data, compute, models, norms) has equal marginal sovereign returns per unit of budget. In simpler terms, early investments give you big jumps; later investments are flat. That is why budget should flow first to pillars with low $x_i$.
\end{itemize}
\begin{itemize}
    \item Sovereignty is best maximized when governments spend in tandem on data and compute infrastructure, as reflected in the complementarity term $\theta$. So, if compute capacity C is already strong, the marginal return on data spending rises, because more data makes the compute capacity usable for models. Merely focusing on model development is not enough.
\end{itemize}
\begin{itemize}
    \item Budget constraint is an important part of the sovereignty equation. If the shadow price of public funds ($\mu$) is high, then only the highest-return pillars should be funded. This is an important conclusion for developing countries where the marginal cost of public funds can be relatively high. For example, Ahmad and Stern (1987) estimated the marginal cost of public funds in India to range between 1.54 and 2.17. Kelkar and Shah (2019) speculate that it may be as high as 3. In general, $\frac{\mu}{\alpha}$ is the bar that marginal returns must clear. If a pillar’s slope at zero doesn’t clear the bar, the planner should assign zero budget there.
\end{itemize}

As a general heuristic for policy makers, they need to fund each pillar of sovereignty until the next rupee of spending produces the same incremental sovereignty gain across all four. They should set openness where the marginal global benefit of collaboration equals the marginal sovereignty cost of dependency. 

\section{Applied case: India’s position and a testable policy rule}

The IndiaAI Mission was approved with a Rs. 10,372 crore outlay to bridge ecosystem gaps and “ensure tech sovereignty,” with pillars on compute, foundation models and “Safe \& Trusted AI.” The compute infrastructure is being built: media briefings from the Mission CEO indicate thousands of GPUs already tendered, with a third tender adding \~3,850 GPUs and 1,050 Google Trillium TPUs (Suraksha P \& Agarwal, 2025). AIRAWAT, the AI supercomputer providing mixed-precision capabilities installed at C-DAC, Pune, featured on the global TOP 500 Global Supercomputing List at the International Supercomputing Conference (ISC 2023) in Germany (Ministry of Electronics \& IT, 2023). This signals progress, but utilization rates and cost per GPU-hour for Indian researchers are not yet regularly published, making it hard to benchmark marginal returns.

Bhashini (National Language Translation Mission) is building open multilingual resources and tooling; the mission documents set out national-scale language data creation and crowdsourcing (Bhasha Daan). Public programs like \textit{Bhasha Daan} demonstrate that curated Indic language data can be mobilized at scale. However, metrics on dataset quality, lineage, and downstream use remain patchy, limiting a precise measure of returns on investment. Future work must develop sovereignty metrics that measure not only dataset counts but usability, provenance, and reusability.

India has not yet released a fully homegrown foundational model, though start-ups like Sarvam AI are piloting Indic LLMs and the Mission calls for government-funded foundation models. Current output is limited to fine-tuned or adapted systems; this is an area where marginal gains are small relative to global leaders. 

India has been proactive in setting a normative agenda— through the Safe \& Trusted AI pillar, the planned IndiaAI Safety Institute, and AIKosha’s Responsible AI tools. While promising, these remain new institutions with few audits completed at scale. 

India participates actively in the Global Partnership on Artificial Intelligence (GPAI) and global standard-setting, while also mandating on-shore data storage and seeking sovereign cloud arrangements. The model suggests India’s current openness level is neither autarkic nor fully globalized.

Based on available evidence, India’s position can be summarized as follows. India has clear sovereign footholds in data and compute, some institutional momentum in norms, but a weaker position on model autonomy. The bigger challenge is integration: investments in GPUs and datasets are not yet tightly coupled, raising the risk of stranded resources.

The model yields three immediate implications for India:

\begin{enumerate}
  \item Use the bar $\mu / \alpha$ to gate spending decisions: $\mu$ (shadow price / marginal cost of public funds) is typically $> 1$ when budgets are tight. In the absence of more recent empirical estimates, we adopt the India-specific values reported by Ahmad and Stern (1987), who estimate the marginal cost of public funds to range between 1.54 and 2.17. If policymakers set $\alpha \approx 0.7$, then the implied bar lies between 2.2 and 3.1. The implication is that programs that deliver little sovereignty per rupees---such as idle GPUs, datasets without downstream use, and audits that fail to cover high-risk systems---should not expand until their adjusted unit costs fall below this bar.

  \item Apply the openness rule now---default to an interior level: In today's India context, with clear chip and cloud dependence but strong benefits from collaboration, the default should be managed openness. India would be better served by keeping research partnerships, open-source participation, standards work, and import of tools/chips. However, it can impose on-shore inference for sensitive public services, escrowed adapters/APIs or exit clauses, and domestic fallbacks for critical workflows.
 
    \item Priorities implied by current gaps: There are certain areas that can be emphasized as the first steps: 
   \begin{enumerate}
    \item Pair Data \& Compute (high complementarity $\theta$): set joint targets so every rupee in Bhashini-style datasets is matched by booked sovereign GPU hours, and vice-versa. This prevents “stranded” capacity. 
    \item Harden ModelOps: stand up continuous pre-deployment cards, update/change logs, drift monitors, and bias/safety audits for public-facing models—make passing these a condition for grants/hosting. 
    \item Choose model scope smartly: prefer fine-tuned domain models for priority services (such as health, agriculture and welfare programs) over chasing a single frontier LLM. This will help clear the $\mu$/$\alpha$ bar more often.
    \item  Procure openness with safeguards: when buying cloud/API capabilities, write in data residency, telemetry, migration/exit terms, and staged access—so exposure cost stays low relative to benefit.

\end{enumerate}
\end{enumerate}

As a pilotable mechanism to make the model testable, further operationalization of the model would involve creating a quarterly marginal returns dashboard. For example, weights can be measured through light, transparent policy scoring (ministries, states, academia, civil society) and then normalized to $\sum w_i = 1$ = 1. Though it may be politically sensitive, the process is methodologically straightforward---structured scoring methods like pairwise comparisons or Analytic Hierarchy Process can ensure transparency and repeatability. Openness can be converted into a checklist (so $\theta^*$ is a living choice). Incremental benefit can be measured as the benefit from last notch of openness, such as speed-to-deploy and CAPEX avoided, while incremental exposure can be measured via lock-in cost to exit/migrate/retrain, sensitivity of data, and cross-border compliance risk. Partnerships and cloud decision can be approved only if benefit $\geq$ exposure for that increment.

Further the country can institutionalize two guardrails:
\begin{itemize}
    \item Joint $D \times C$ OKR: require target GPU utilization and dataset-backed bookings (for example, more than 75\% utilization and more than 40\% of booked hours tied to Indic datasets).
\end{itemize}
\begin{itemize}
    \item ModelOps gates: deployment of public-facing models can be linked to pre-deployment cards, change logs for updates/RLHF, and bias/safety audits for high-risk use. All grants and hosting can be linked to these gates.
\end{itemize}

\section{\textbf{Applied case: Middle East’s state‑led pathway to sovereign AI} }

Saudi Arabia and the United Arab Emirates offer a distinctive, state‑led pathway to sovereign AI. Large public investment in Arabic-first models and sovereign cloud imply a high policy weight on sovereignty ($\alpha$), lower shadow price of funds than in fiscally constrained settings (lower effective $\mu$), and strong complementarities between data and compute (high $\theta$). The key policy concern is to set openness ($O$) so that global collaboration (chips, clouds, models) is harnessed with guardrails, rather than shunned. Where India faces tighter fiscal constraints (higher $\mu$) and must ration compute, the Gulf’s model shows what high‑$\alpha$, lower‑$\mu$, high‑$\theta$ strategies can achieve provided ModelOps and normative alignment scale as quickly as hardware. The core lesson holds across both contexts: pair data with compute, treat openness as a tunable policy variable, and embed lifecycle governance so that sovereign inputs produce sovereign outcomes.

Rather than autarky, both Saudi Arabia and the UAE are moving toward managed interdependence. The G42--Microsoft deal in the UAE (Microsoft, 2024) includes an inter-governmental assurance agreement and AI safety and security commitment, which reduce exposure risk $\lambda P(O)$ while preserving global benefits $(1 - \alpha) G(O)$. Abu Dhabi's sovereign cloud architecture with Core42 and Microsoft aims to keep governance and residency local while enabling access to global innovations, model catalogs, and toolchains---an archetypal interior solution for openness. The UAE's federal data protection law (Decree-Law 45/2021) establishes national-level governance and cross-border provisions, while Arabic-first LLMs Falcon-180B (TII) and Jais-13B (G42/MBZUAI/Cerebras) represent an open-source track that balances self-reliance with global collaboration. 

In Saudi Arabia, deploying ALLaM (an Arabic LLM reported to rank high on Arabic MMLU benchmarks) on DEEM Cloud alongside commercial tooling (e.g., IBM watsonx) is a similar hybrid: local control over inference and data, global access where safe and useful. Saudi Arabia’s DEEM Government Cloud operated by SDAIA’s National Information Center provides IaaS/PaaS/SaaS for public entities and is explicitly positioned to support digital sovereignty and cost efficiency. SDAIA reports multi‑billion riyal fiscal savings from DEEM adoption (Saudi Press Agency, 2024). The policy logic for Arabic‑first models is clear: Arabic is used daily by 400+ million people across 20+ countries, and sovereign service delivery requires dialectal and cultural alignment that generic English‑centric models struggle to deliver, especially in delivering public services to the local population.

Converting these broad principles into practice requires focus on three priorities, with measurable targets and policy-ready checklists. 

\begin{itemize}
    \item As a first step, they can set measurable capacity and complementarity targets to exploit the Data x Compute complementarity $\theta$ and clear the spending bar $\mu$/$\alpha$. For example, they need to focus on data usability, not just residency, which may require publishing dataset cards (like provenance and consent) for all government AI datasets and stand-up Arabic dialect data trusts. Quarterly GPU utilization and riyal per GPU‑hour on sovereign clusters can help track utilization and gain cost transparency. Co-locating AI halls with renewables/advanced cooling and scheduling training in off‑peak windows can help push sustainability and low-carbon compute. These can be converted into specific targets for the next 1-2 years, such as more than 75\% sovereign GPU utilization, 40\% of booked hours tied to Arabic fine‑tunes/evals, more than 70\% of public AI datasets from verifiable sources and at least 50\% of compute power from low-carbon energy sources. 
\end{itemize}
\begin{itemize}
    \item To choose an interior $O^*$ and lower exposure $\lambda p$, they will need to engineer managed openness with guardrails, such as assurance agreements in all strategic cloud/API deals, and adopt a benefit–risk scorecard so that each incremental increase in openness is approved only if incremental benefit is higher than the exposure risk. 
\end{itemize}
\begin{itemize}
    \item Another area of focus must be on keeping $N$ high and preventing post-deployment erosion of control. This can be translated into specific targets around annual audit of high-risk systems, number of domain models deployed every year (e.g., related to justice or health), and drift reports for critical services. For example, targets may include 80 percent of high‑risk systems audited annually, initial public/regulatory notification within a certain time limit of any material AI incident, and the number of domain models deployed every year.
\end{itemize}

In general, the focus in Middle East should be on ensuring that sovereign inputs reliably become sovereign outcomes. 

\section{ Conclusion}

Artificial intelligence presents states with the paradox of sovereignty in an interconnected age. On the one hand, national governments increasingly see AI as a strategic resource requiring sovereign control. On the other, the technology’s foundations—globally sourced data, semiconductor supply chains, open-source collaboration, and cross-border standards—resist enclosure. This paper has argued that AI sovereignty should not be seen as a binary condition but as a continuum, defined by the trade-offs between autonomy and interdependence.

By revisiting political theories of sovereignty, examining historical analogies with electricity and the internet, and formalizing the problem as a planner’s allocation under budget constraints, we identified two rules of thumb: equalize marginal returns across the four sovereignty pillars (data, compute, models, norms), and set openness at the point where the marginal global benefit of collaboration equals the marginal cost of dependency. 

Applied to India, these heuristics highlight the need to avoid stranded investments by pairing data with compute, institutionalizing lifecycle governance through ModelOps, and pursuing domain-specific models aligned with public priorities rather than chasing a frontier LLM. India already has sovereign footholds in data and compute, some institutional momentum in norms, but a weaker position in model autonomy. Its challenge is integration. At the same time, India’s participation in global standard-setting—through GPAI and other fora—shows that sovereignty can be exercised through norm entrepreneurship. Using the bar $\mu$/$\alpha$, informed by India-specific estimates of the marginal cost of public funds, provides a transparent threshold for deciding which programs to expand and which to defer. 

Applied to the Middle East (Saudi Arabia and the UAE), a state‑led path with Arabic‑first models and sovereign cloud reflects high sovereignty weights and strong Data×Compute complementarities. Here, managed interdependence translates into openness with enforceable guardrails and exits.

Overall, building a simple quarterly dashboard to track marginal sovereignty returns, and an openness checklist to evaluate partnerships, would allow the sovereign AI model to be empirically tested and iteratively refined.

Across both cases, the lesson is clear: sovereignty in AI will not come from autarky, but from managed interdependence. The strategic task is the same: embedding sovereignty not only in infrastructure, but also in institutions and norms, enabled by measurable capacity and complementarity targets, contractual and technical guardrails on openness, continuous governance of model behavior and alignment, and the ability to negotiate global standards. Sovereign AI, in this sense, is less about closing borders and more about creating the capacity to choose, adapt, and influence within a globally networked AI ecosystem.

 \textbf{Biographical Note}

Shalabh Kumar Singh is principal director for technology thought leadership at Accenture. He has co-authored two books: \textit{Social Accounting Matrix for India: Concepts, Construction and Application} (Sage) has been a recommended reading for students of economics in various institutions in India, and \textit{The Public Innovator's Playbook: Nurturing Bold Ideas in Government} (Harvard's Ash Institute for Democratic Governance and Innovation) was recognized as a key resource by various government organizations across the globe. He has also co-authored several reports and articles in leading journals, including Harvard Business Review, MIT Sloan Management Review, Journal of Business Strategy and Journal of International Development.

Shubhashis Sengupta is a Managing Director and Lead, Accenture Labs India and a Fellow of Accenture Labs. His area of specialization is around large-scale distributed systems (Cloud and Edge), Software Engineering, and Generative AI and AI-based (NLP and Machine Learning) systems. At Accenture, his team has developed several accelerators around Software Engineering Automation, Conversational AI, Knowledge graphs and Text Analytics. He has chartered research agendas for AI and Software Engineering groups. Shubhashis collaborates extensively with leading industry think tanks and academia. Shubhashis has nearly 80 journal and peer-reviewed conference papers, 65+ granted US, Japan, EU, China, and India patents, and numerous applications to his credit. He is a Senior Member of ACM and a Senior Member of IEEE Computer Society. He is a founding member of the AAAI India chapter. He has been on the program committee and review board of several conferences and journals including IEEE Spectrum and IEEE Transactions, Association of Computational Linguistics etc.

\makeatletter
\renewcommand{\@biblabel}[1]{} 
\makeatother


\end{document}